\documentclass{article}

\usepackage{graphicx}

\usepackage[utf8]{inputenc}
\usepackage[T1]{fontenc}
\usepackage{pslatex}
\usepackage{geometry}
\usepackage{authblk}
\usepackage{amsmath}
\usepackage{amssymb}
\usepackage{hyperref}
\usepackage{gensymb}
\newcommand{\etal}{\textit{et al.}}

\usepackage{fancyhdr}
\title{Wind LiDAR measurement campaign at CESAR Observatory in Cabauw: preliminary results}
\author{Steven Knoop\thanks{steven.knoop@knmi.nl}} 
\author{Willem Koetse}
\author{Fred Bosveld}

\affil{Royal Netherlands Meteorological Institute (KNMI), P.O. Box 201, 3730 AE De Bilt, The Netherlands}

\date{\today}

\begin{document}  

\maketitle
	
\begin{abstract}
A two-year verification campaign of the ZephIR~300 vertical profiling wind LiDAR has recently been started by the Royal Netherlands Meteorological Institute (KNMI) at the CESAR Observatory (Cabauw Experimental Site for Atmospheric Research) in Cabauw, The Netherlands. In this paper we present preliminary results of the first six months of this campaign (February 14 - July 31, 2018), focusing on the (height-dependent) data availability of the ZephIR~300 under various meteorological conditions (precipitation intensity, cloud base height and visibility) and the data quality via a comparison with in situ wind measurements at several levels in the 213-m tall meteorological mast. 
\end{abstract}
	
\section{Introduction}

The ZephIR~300\footnote{More precisely, the offshore version ZephIR~300M, which however has the same (functional) specifications and output as the ZephIR~300. Therefore, we refer to the instrument as ZephIR~300 throughout this paper.} wind LiDAR (Light Detection And Ranging) has recently been selected to be installed on offshore substations within the upcoming offshore wind farms in the Dutch North Sea (Borssele, Hollandse Kust Zuid, Hollandse Kust Noord, to be commissioned in 2019, 2020 and 2021, respectively). Its main purpose is to continuously measure the wind speed and direction at hub height, such that in the event of an unavailable offshore electricity grid compensation can be determined on basis of the actual wind conditions \cite{schaderegeling}. The wind LiDAR data will be made freely available via the Royal Netherlands Meteorological Institute (KNMI).

In this context the KNMI has started February 2018 a two-year measurement campaign of this particular ZephIR~300 instrument at the CESAR Observatory (Cabauw Experimental Site for Atmospheric Research) in Cabauw, The Netherlands \cite{monna2013,CESAR}, in order to obtain detailed information about the data availability and quality, under various meteorological conditions and for the full height range of the instrument. A comparison between the wind LiDAR and in situ wind measurements at several levels in the 213~m high meteorological mast is made, while monitoring the meteorological conditions with several collocated in situ and remote sensing instruments, such as visibility and present-weather sensors, rain gauges, and ceilometers. 

In this paper we present preliminary results of the first six months (February 14 - July 31, 2018) of this measurement campaign. First, a short introduction is given of the ZephIR~300 instrument (Sect.~\ref{ZephIR intro}) and the CESAR Observatory (Sect.~\ref{Cabauw intro}). Secondly, the measurement campaign is further outlined (Sect.~\ref{Measurement campaign}), including the configuration of the ZephIR~300, after which the results on data availability (in correlation with precipitation intensity, first cloud base height and visibility) (Sect.~\ref{da}) and data quality (horizontal wind speed and wind direction) (Sect.~\ref{dq}) are presented. Finally, conclusions are drawn and an outlook is given (Sect.~\ref{conclusions}).

\begin{figure}
  \centering
  \includegraphics[width=\textwidth]{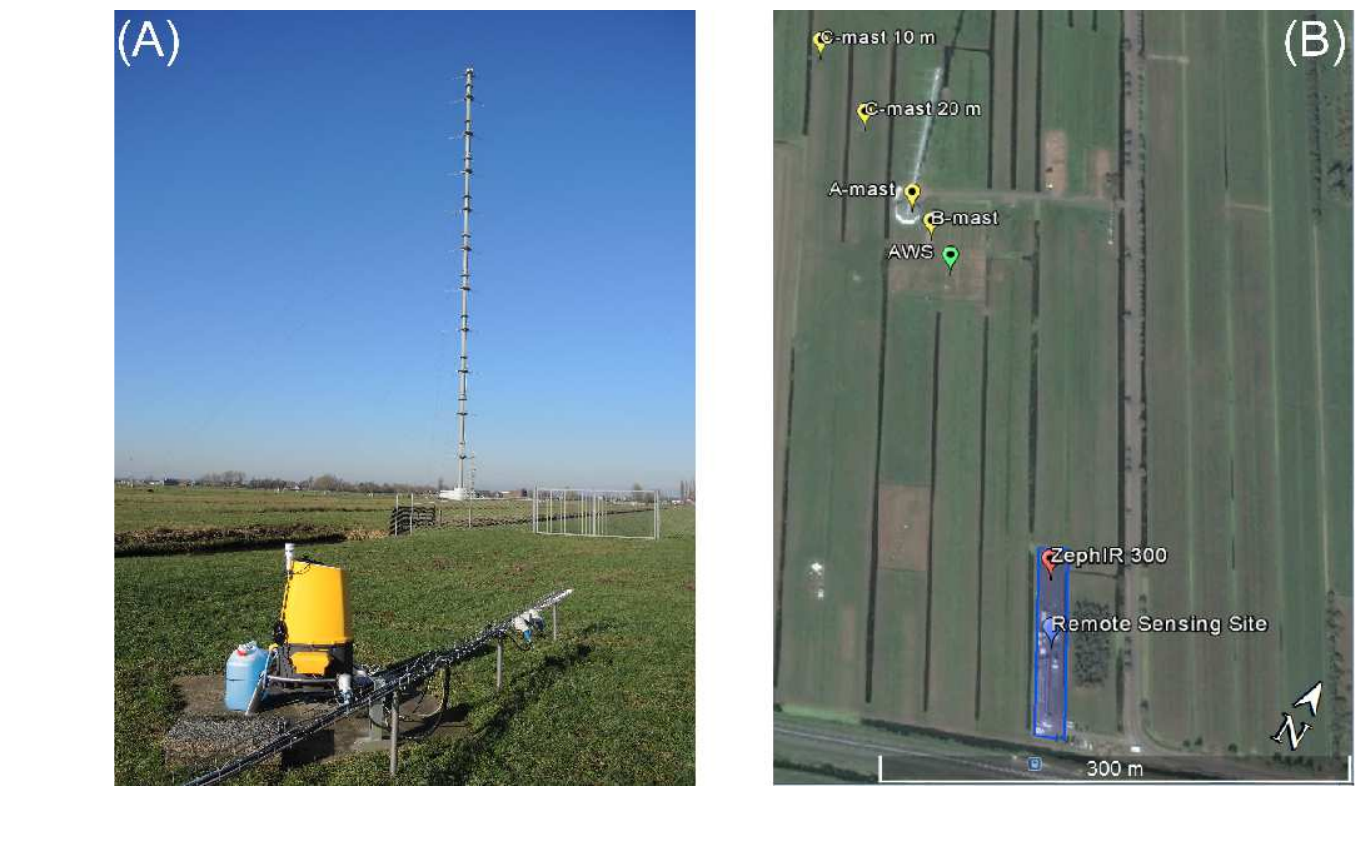}
  \caption{(A) Photo of the ZephIR~300 instrument at CESAR Observatory in Cabauw, the Netherlands (location 51.971\degree N, 4.927\degree E), with the 213-m tall A-mast visible the background (view in NW direction). (B) Overview of (part of) the CESAR Observatory, indicating the locations of the A-, B- and C-masts (yellow), automatic weather station (AWS) (green), remote sensing site (blue), and ZephIR~300 (red).}
  \label{ZephIR@Cabauw}
\end{figure}

\section{ZephIR~300 wind LiDAR}\label{ZephIR intro}

The ZephIR~300 (ZephIR Lidar, UK) is a vertical profiling wind LiDAR, providing the wind speed and direction up to a few hundred meters above the instrument by measuring the Doppler-shift of the backscattered laser light, typically by aerosols in the moving air. More specifically, the ZephIR~300 is a homodyne, continuous-wave (CW) focusing wind LiDAR, in which the laser beam is transmitted through a constantly rotating prism (wedge) to perform a so-called velocity azimuth display (VAD) scan, with a scanning cone angle of 30\degree, in order to be sensitive to the horizontal wind speed. For each height one complete rotation takes 1~s, in which 50 measurements of 20~ms are taken, from which the 3D wind vector is reconstructed (i.\,e.\,horizontal and vertical wind speed, horizontal wind direction). The manufacturer specifies the wind speed and wind direction accuracies as <0.1~m/s and <0.5\degree, respectively, and the height range is 10-200~m above the instrument\footnote{The height of the instrument is 1~m.}, although up to 300~m can be selected in the software. There is a maximum of 10 user-configurable measuring heights, besides a pre-fixed height of 38~m above the instrument, which all are measured sequentially by changing the focus of the laser beam after each VAD scan. 

Being a CW focusing wind LiDAR, the probe length increases quadratically with height: at 10~m height above the instrument the probe length is 0.07~m, whereas at 200~m it is 30~m. Furthermore, the ZephIR~300 is sensitive to clouds that are above the maximum range, as the contribution to the Doppler signal from clouds in the tail of the laser pulse profile can be comparable to the aerosol signal at the preselected focusing height \cite{smith2006wle}. A cloud removal algorithm is used to correct for this effect, which involved a measurement at an additional greater height \cite{kindler2007aem,courtney2008tac}. As a result of the homodyne detection, meaning that only the absolute value of the Doppler-shift is measured, there is a 180\degree~ambiguity in the measured wind direction. To solve this issue the ZephIR~300 includes an attached met station (AIRMAR WeatherStation 200WX), which contains a sonic anemometer to measure the wind direction just above the instrument and provides an estimate for the remotely measured wind direction \cite{courtney2008tac}. Still, a few percent of incorrectly assigned wind direction events is possible (see e.\,g.\,Ref.~\cite{wouters2016vot}). A more extensive introduction of the ZephIR~300 (and its predecessors) is given in Ref.~\cite{pitter2015itc}.

The instrument outputs, besides raw data, quality controlled 10-minutes averaged data, including horizontal and vertical wind speed, horizontal wind direction, minimum, maximum and standard deviation of the horizontal wind speed, and turbulence intensity. The analysis in this paper is based on this 10-minutes averaged data. 

\section{CESAR Observatory}\label{Cabauw intro}

The measurement campaign takes place at the CESAR Observatory in Cabauw, The Netherlands \cite{CESAR}, which is centered around a 213-m tall meteorological mast ("A-mast"), see Fig.~\ref{ZephIR@Cabauw}. Its instrumentation and siting, in particular in the context of wind measurements, have been extensively described \cite{ulden1996abl,verkaik2007wpm}. Wind speed and wind direction are measured with cup anemometers and wind vanes, respectively, at six levels: 10, 20, 40, 80, 140 and 200~m. Precautions are taken to avoid too large flow obstruction from the A-mast and the main building at the bottom of the A-mast. At the levels 40, 80, 140, and 200~m of the A-mast the wind direction is measured at three booms and wind speed is measured at two booms. At the levels 10 and 20~m the wind direction and wind speed are measured at two separate, smaller masts south ("B-mast", 30~m SE from A-mast) and north (two "C-masts", 70~m and 140~m NE from A-mast for 20~m and 10~m level, respectively) of the main building; the selection between these two masts depends on the wind direction. The data is quality controlled, including corrections for flow distortions from the mast.

For monitoring the meteorological conditions a KNMI automatic weather station is operated 100~m SE of the A-mast. In particular, this includes a KNMI rain gauge for precipitation intensity, a Vaisala FD12P present-weather sensor for precipitation type and visibility (at a height of 2~m above ground level) and a Lufft CHM15K ceilometer, which also outputs the first cloud base height. In addition, Biral SWS-100 present-weather and visibility sensors are located in the A-mast (at 40, 80, 140 and 200~m) and in the B-mast (at 2, 10 and 20~m).

\begin{figure}
  \centering
  \includegraphics[width=\textwidth]{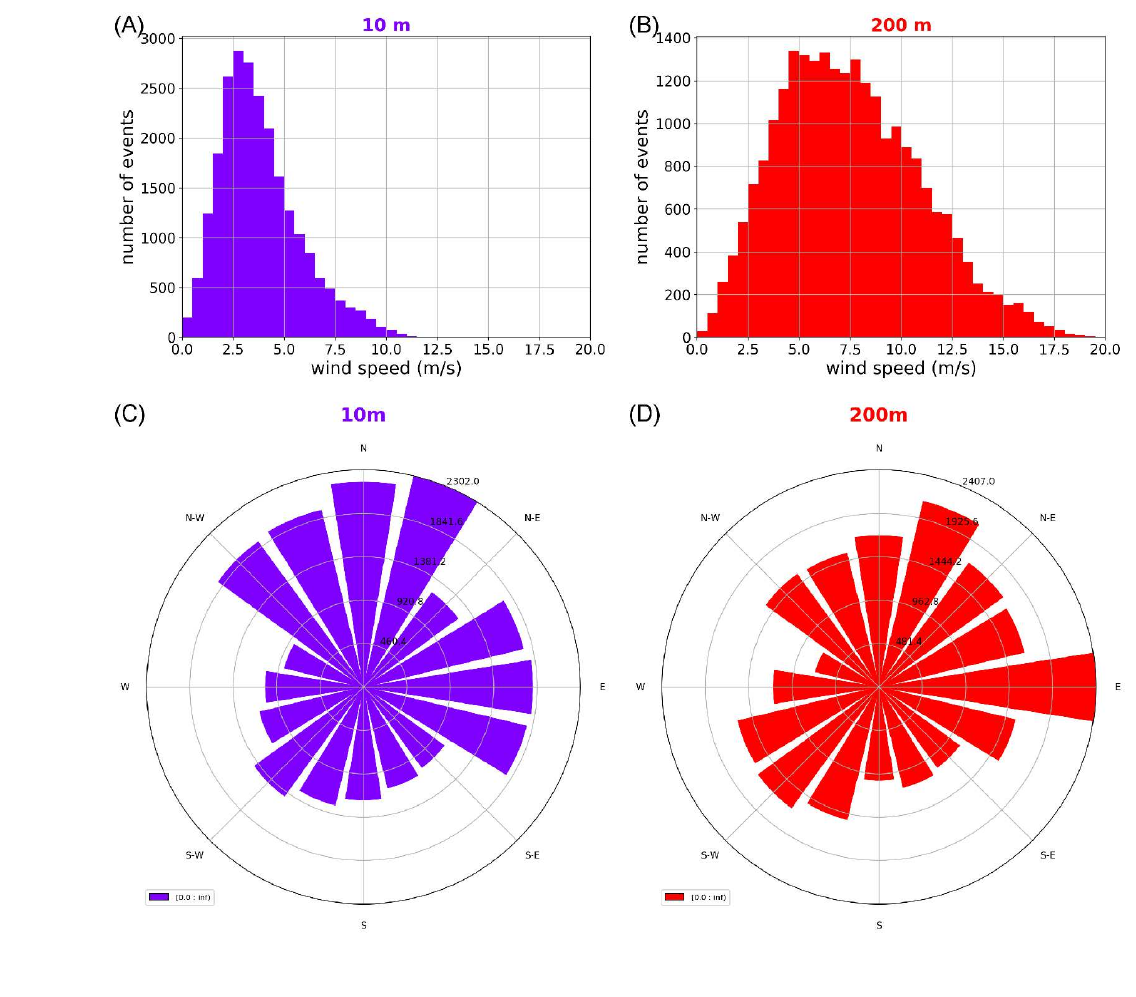}
  \caption{Overview reference wind conditions at the CESAR Observatory in period February 14 until July 31, 2018: (A), (B) wind speed histograms (bin size 0.5~m/s) and (C), (D) wind roses (bin size 22.5\degree) for a height of 10 and 200~m, respectively.}
  \label{Cesar_histogram_windrose}
\end{figure}

\section{Measurement campaign}\label{Measurement campaign}

The ZephIR~300 instrument is placed at the northern part of the remote sensing site (RSS) of the CESAR Observatory, 290~m in SE direction from the A-mast. The instrument is configured to measure at seven heights: 11, 20, 39, 80, 140, 200, 252~m above ground level (agl), matching the levels of the reference wind measurements as close as possible. Note the minimum range of the ZephIR~300 is 10~m above the instrument, corresponding to a minimum height of 11~m, which is compared with 10~m reference level. Also, the pre-fixed height of 39~m (agl) is compared with the 40~m reference level\footnote{The ZephIR~300 configuration requires 5~m separation between the distinct heights, which does not allow to select 40~m because of the pre-fixed 39~m.}, however, this height difference might be irrelevant as the probe length at this height is about 1~m. While the specified range of the ZephIR~300 is limited to 200~m, heights up to 300~m can be chosen in the software. Here we have added 252~m (agl) to the ZephIR~300 configuration, which of course cannot be compared with wind measurements of the mast, but which overlaps with one of the lower levels of a collocated radar wind profiler. However, in this paper we will only consider the data availability of this particular height. 

Note that the distances between the reference masts and the ZephIR~300 instrument are relatively large (see Fig.~\ref{ZephIR@Cabauw}), compared to most other validation or verification studies, meaning that the correlation between the corresponding measurements might be limited, especially for the lower heights. Furthermore, for the two lowest reference levels (10 and 20~m) the distance between the reference and ZephIR~300 depends on the wind direction, the B-mast being much closer than the two C-masts. At those levels, the constructions on the RSS (portacabin, radars) and trees on the east side of RSS, might also affect the correlation between mast and ZephIR~300 (see Sect.~\ref{dq}). 

The final alignment and leveling of the ZephIR~300 at RSS took place on February 14, 2018, marking the start of the measurement campaign. In this paper we consider the period up to July 31, 2018. An overview of the reference wind conditions in this period, at a height of 10 and 200~m, are shown in Fig.~\ref{Cesar_histogram_windrose}.

\begin{figure}
  \centering
  \includegraphics[width=0.5\textwidth]{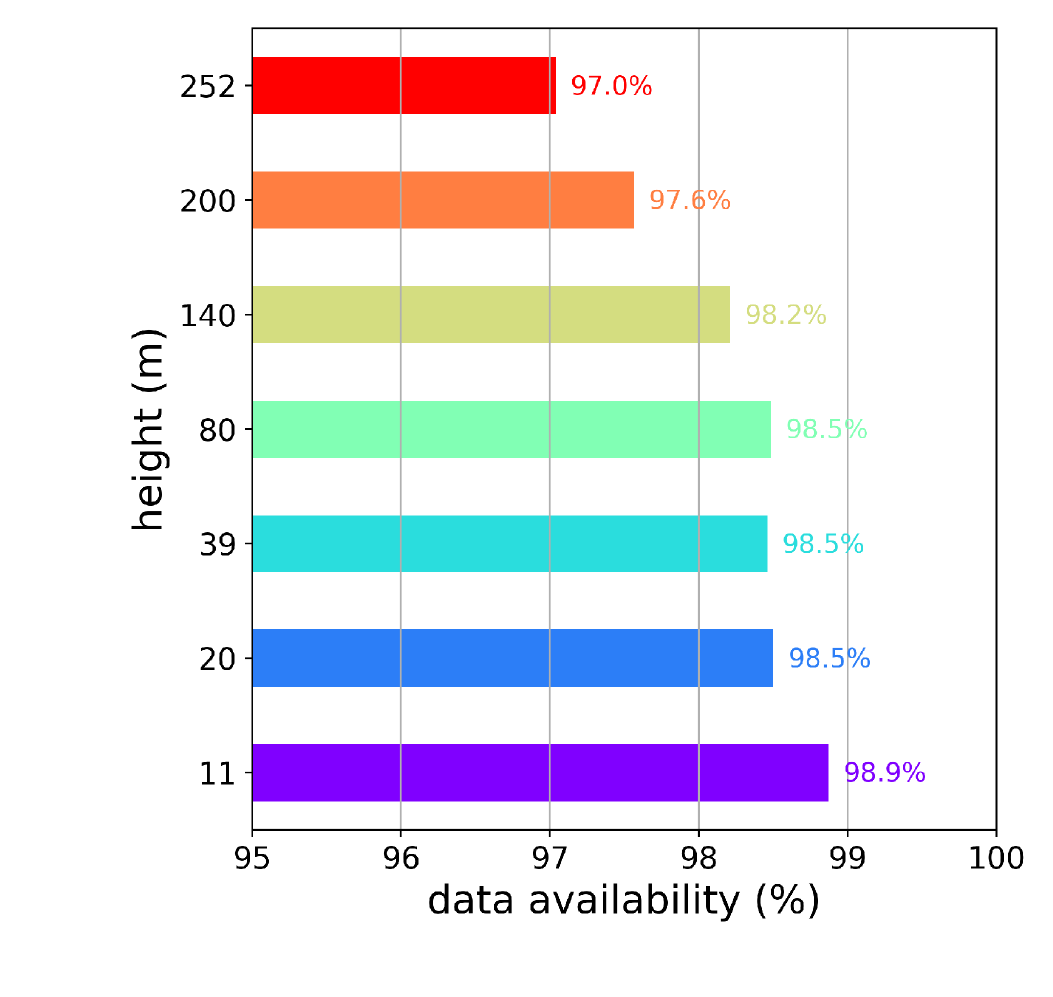}
  \caption{Overall data availability of the ZephIR~300 for the different heights.}
  \label{fig_da}
\end{figure}

\section{Preliminary results}

\subsection{Data availability}\label{da}

\begin{figure}
  \centering
  \includegraphics[width=0.82\textwidth]{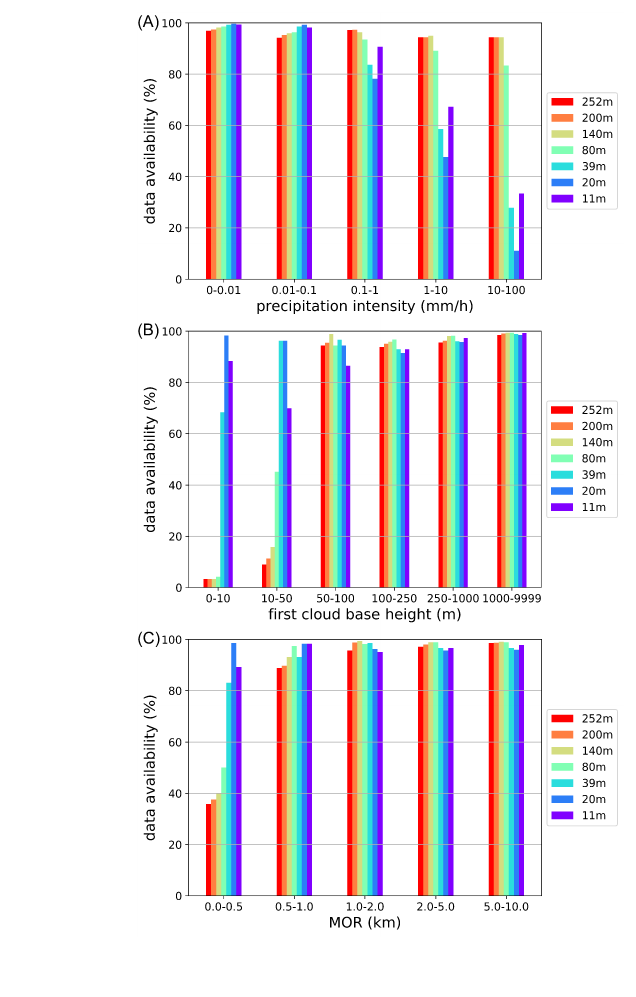}
  \caption{Data availability of the ZephIR~300 for different classes of (A) precipitation intensity, (B) first cloud base height and (C) visibility in terms of MOR (meteorological optical range) at 2~m above ground level.}
  \label{fig_da_conditions}
\end{figure}

The overall data availability for the different heights are shown in Fig.~\ref{fig_da}. Here data availability is defined as the percentage of events the ZephIR~300 outputs quality controlled data, during the time the instrument is switched on (i.e. not considering periods the power is switched off). ZephIR~300 indicates bad events when detecting partial obscuration of the ZephIR window or significant interference with the laser beam at the specified height, or atmospheric conditions which adversely affect LiDAR wind-speed measurements (e.\,g.\, thick fog). In general we find a high percentage of data availability, ranging between 97\% and 99\%, which shows a decreasing trend with height.

In Fig.~\ref{fig_da_conditions} the data availability is further specified for different meteorological conditions. In Fig.~\ref{fig_da_conditions}(A) the data availability for the different heights is shown for different precipitation intensity classes. Here we notice a clear impact on the lower heights for heavy precipitation, whereas the higher heights are hardly affected. This is related to the height-dependent probe length: where at low heights (short probe lengths) individual hydrometers can have a detrimental impact on the data, at high heights (long probe lengths), individual hydrometers are not resolved.

In Fig.~\ref{fig_da_conditions}(B) the data availability for the different heights is shown for different first cloud base height classes, as derived from the ceilometer data. Down to a cloud base height of 50~m there is only a small decreasing trend, whereas below 50~m the data availability for the higher heights drops dramatically. A similar trend is seen in Fig.~\ref{fig_da_conditions}(C), in which the data availability for different visibility classes is shown (in terms of meteorological optical range (MOR)), where visibility is measured at 2~m above ground level. For MOR>1~km visibility has no impact on the data, but below 1~km (fog conditions) the data availability of the higher heights drops, most notably below 0.5~km. Thus low clouds and fog can limit the amount of data, especially the higher heights, however the overall impact here is still quite small, as can be seen from Fig.~\ref{fig_da}. 

\subsection{Data quality}\label{dq}

\begin{figure}
  \centering
  \includegraphics[width=\textwidth]{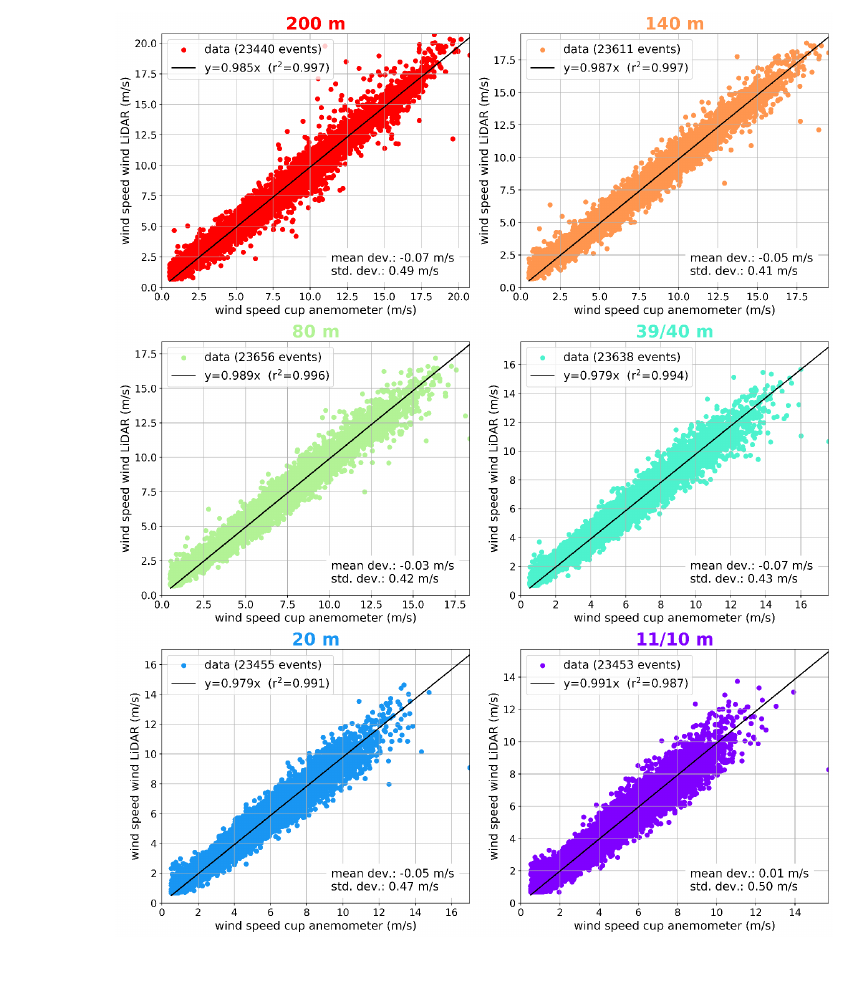}
  \caption{Wind speed comparison between ZephIR~300 and reference mast data for the different heights, in which the results of a linear regression analysis (slope and correlation coefficient), and the mean deviation and its standard deviation are also shown.}
  \label{scatterplot_windspeed}
\end{figure}

Fig.~\ref{scatterplot_windspeed} shows the correlation between the ZephIR~300 and the reference mast wind speed measurements (cup anemometer), where a linear regression analysis provides the slope ($y={\rm slope}\times x$) and the correlation coefficient $r^2$. Also the mean deviation and the standard deviation in the deviation is given. The data is only filtered on the condition that the cup anemometer (for the specific height) reports a value larger than 0.5~m/s. As explained above, the ZephIR~300 data at 11~m and 39~m are compared with the reference measurements at 10~m and 40~m, respectively. Overall the correlation is very good, with slope and correlation coefficient close to 1 (>0.98 and >0.99, respectively). The mean deviation is below 0.1~m/s, while its standard deviation is 0.4-0.5~m/s.

\begin{figure}
  \centering
  \includegraphics[width=\textwidth]{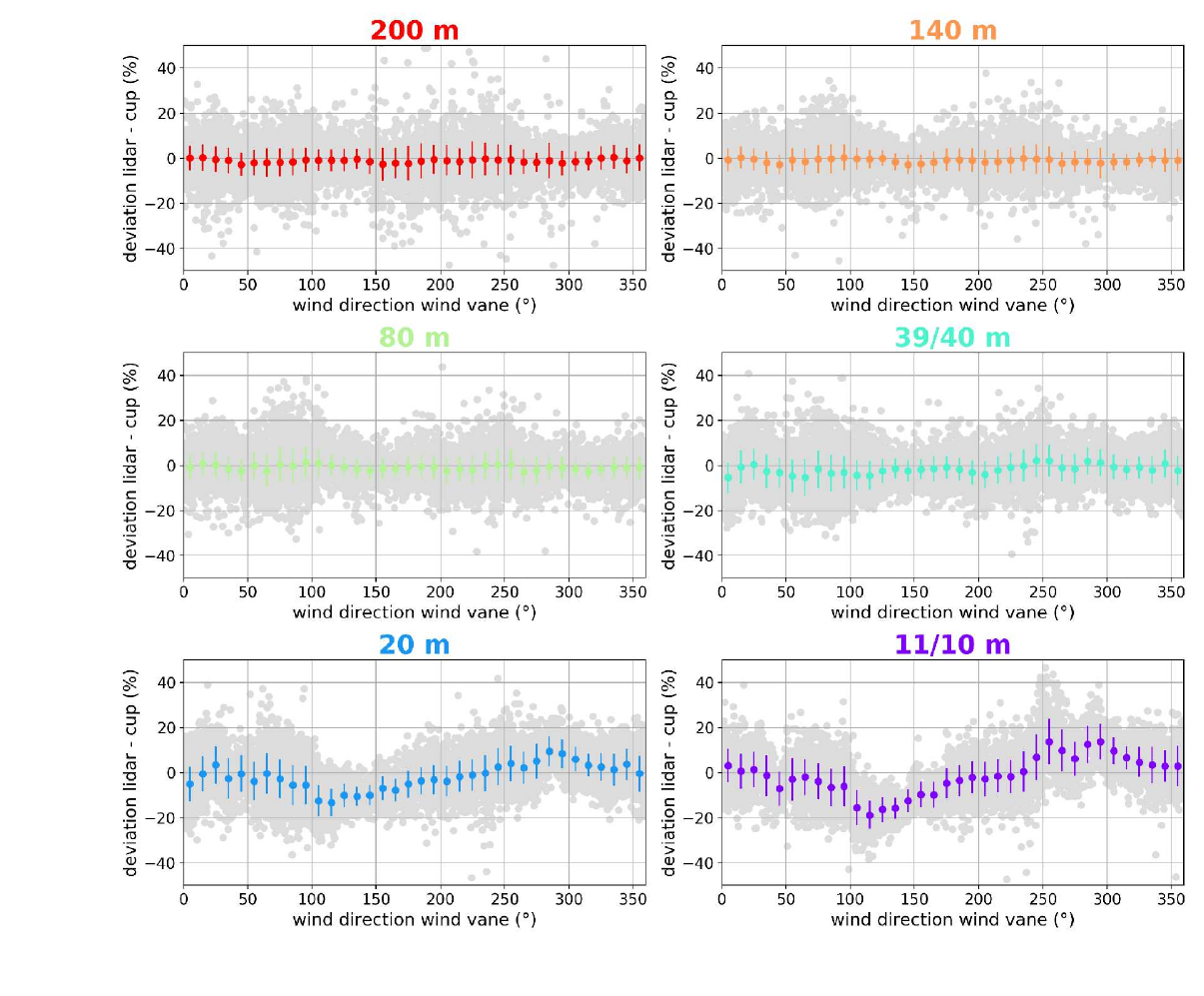}
  \caption{Relative deviation between the the ZephIR~300 wind LiDAR and the reference mast wind speed measurements, where the colored data points represent the bin average and standard deviation (bin size is 10\degree) and the gray data points the unbinned data. The data is filtered on the condition that the cup anemometer (for the specific height) reports a value larger than 4~m/s.}
  \label{winddirectioneffect}
\end{figure}

In Fig.~\ref{winddirectioneffect} the relative deviation between the ZephIR~300 and the reference mast wind speed measurements is shown as function of the reference wind direction, in which the data is collected in bins of 10\degree~and filtered on the condition that the cup anemometer (for the specific height) reports a value larger than 4~m/s. For the upper four heights (for which reference data is measured in the A-mast), no wind direction dependence of observed, whereas for the lower two heights (reference is either B- or C-masts) a clear modulation of the relative deviation is visible, with a negative deviation between 100\degree~and 150\degree~and positive deviation between 250\degree~and 300\degree. We explain this behavior by the presence of flow obstruction at the remote sensing site and neighboring trees SE of the ZephIR~300 (for the 100\degree-150\degree~sector), and neighboring trees NW of the C-masts (for the 250\degree-300\degree~sector), in combination with the relatively large distance between the ZephIR~300 and the reference masts. These sectors will be omitted in future analysis for those heights.

\begin{figure}
  \centering
  \includegraphics[width=\textwidth]{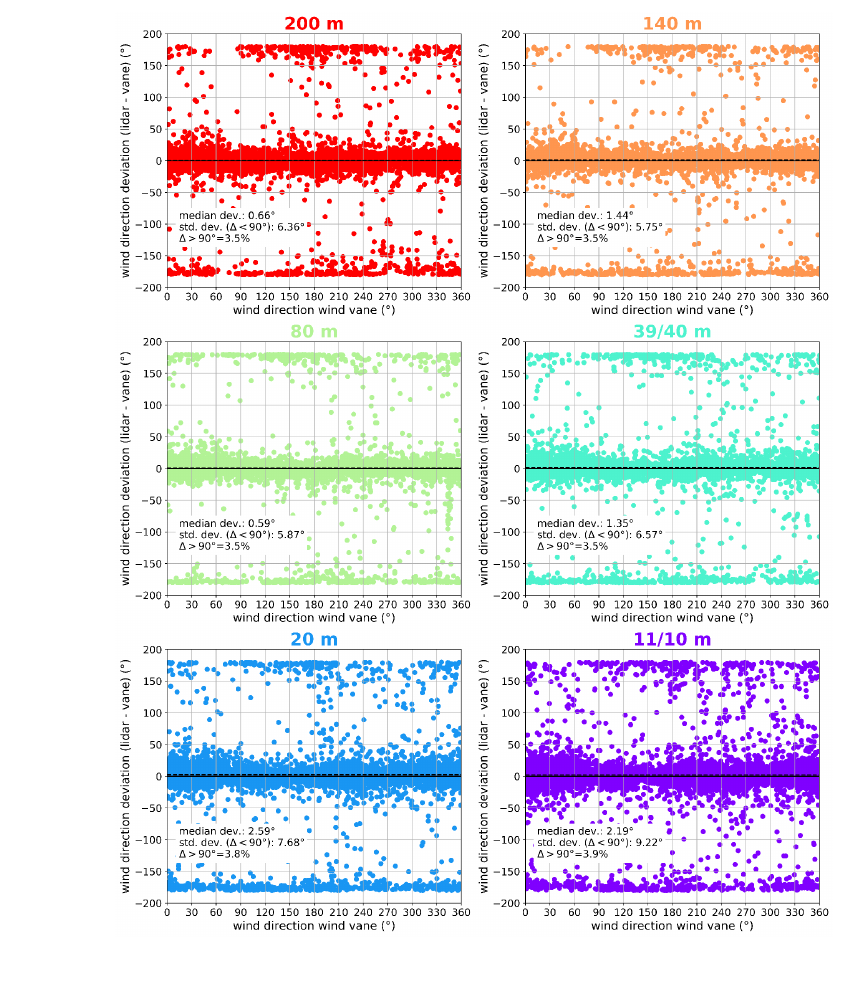}
  \caption{Wind direction comparison between ZephIR~300 wind LiDAR and reference mast data for the different heights, plotting the deviation between the two against the reference mast data. The median deviation (dashed line) and the standard deviation of the data for which the absolute value of deviation is less than 90\degree~($\Delta<90\degree$) is given, as well as the percentage of events for which the absolute value of the deviation is larger than 90\degree~($\Delta>90\degree$).}
  \label{scatterplot_winddirection}
\end{figure}

Fig.~\ref{scatterplot_winddirection} shows the correlation between the ZephIR~300 and the reference mast wind direction measurements (wind vane), plotting the deviation of the ZephIR~300 from the wind vane against the wind vane data. The median and the standard deviation of the data, for which the absolute value of the deviation is less than 90\degree, is given. Again, the data is filtered on the condition that the cup anemometer (for the specific height) reports a value larger than 0.5~m/s, and the ZephIR~300 data at 11~m and 39~m are compared with the reference measurements at 10~m and 40~m, respectively. The median deviation is below 3\degree (for the upper four heights less than 1.5\degree), and its standard deviation below 10\degree.

\begin{figure}
  \centering
  \includegraphics[width=\textwidth]{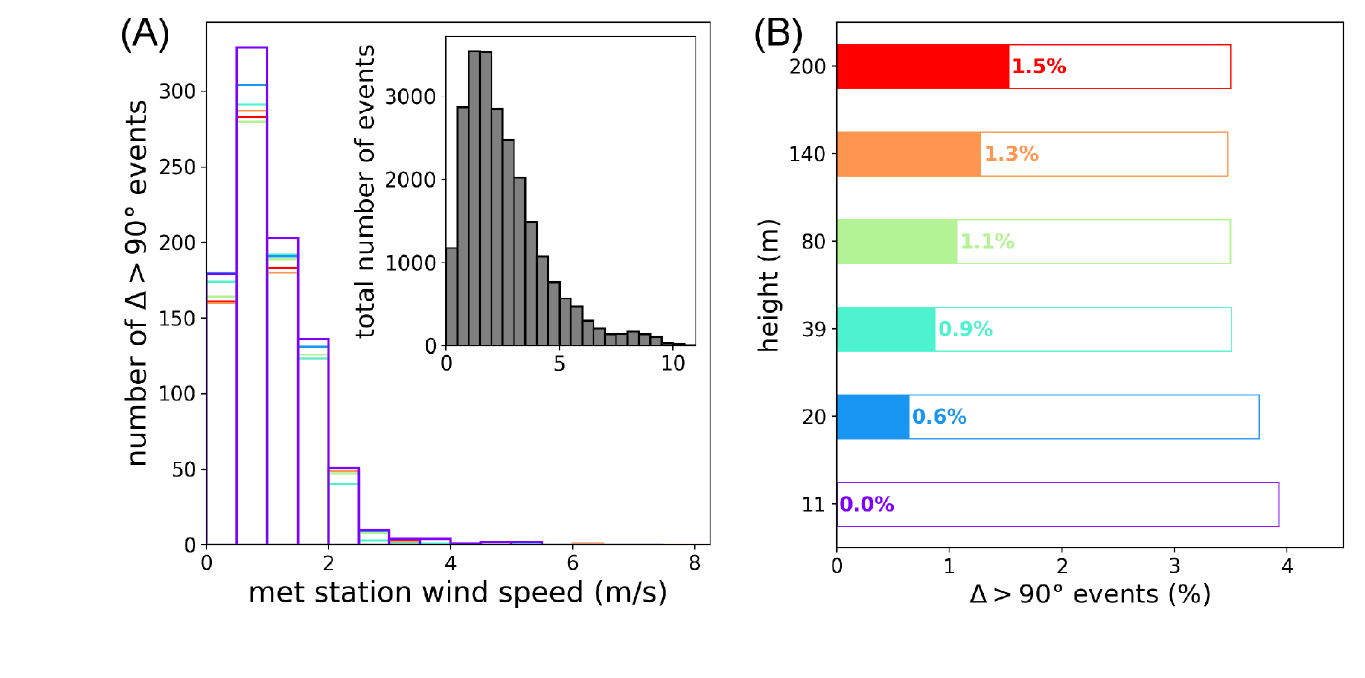}
  \caption{(A) Histogram of the reported wind speeds (bin size 0.5~m/s) of the met station for $\Delta>90\degree$ events for the different heights (color coding as in the other figures), where the inset shows the histogram for all events, demonstrating that these $\Delta>90\degree$ events correlate almost exclusively with very low wind conditions. (B) Percentages of $\Delta>90\degree$ events for different heights, open bars as given by the instrument, filled bars after applying a correction scheme based on the 10-m reference wind direction, resulting in a significant reduction of the 180\degree~error.}
  \label{180error}
\end{figure}

Some part of the data points in Fig.~\ref{scatterplot_winddirection} lies around $\pm$180\degree~deviation, which is due to the 180\degree~ambiguity issue of the ZephIR~300 as mentioned above. We parametrize this effect by considering the number of events for which the absolute value of the deviation is larger than 90\degree, indicated as "$\Delta>90\degree$". We find that the percentage of $\Delta>90\degree$ is similar for the different heights and lies between 3.5\% and 3.9\%. These events almost exclusively (more than 97\%) occur in very low wind conditions, in which the met station on top of the ZephIR~300 (at 1-m height agl) reports a wind speed of less than 2.5~m/s, as shown in Fig.~\ref{180error}(A).

With additional information about the actual wind direction, one can correct the $\Delta>90\degree$ events. In case such information is only available at a single height, one can compare this with a corresponding (or nearby) height of the ZephIR~300, and for timestamps with $\Delta>90\degree$ for this height change the ZephIR~300 wind direction data for \emph{all} heights by 180\degree. In Fig.~\ref{180error}(B) we show the result of such a correction on basis of the 10-m reference wind data. By definition, this results in 0\% $\Delta>90\degree$ events for the corresponding 11-m ZephIR~300 data, but also a significant reduction of $\Delta>90\degree$ for the other heights, in which $\Delta>90\degree$ increases with height due to the natural wind veer that reduces the correlation between the wind direction of the different heights.

\section{Conclusions and Outlook}\label{conclusions}

We have presented preliminary results of the first six months of our two-year measurement campaign in Cabauw to investigate the data availability and data quality of the ZephIR~300 wind LiDAR. In this paper we have considered only the 10-minute averaged data of the horizontal wind speed and wind direction. The ZephIR~300 is configured to measure at heights of 11, 20, 39, 80, 140, 200 and 252~m above ground level. We have studied the data availability under various meteorological conditions, in particular as function of the precipitation intensity, first cloud base height and visibility. We find an overall data availability of 97\% or better, where a precipitation intensity more than 1~mm/h reduces the amount of data for the lower heights, and low clouds (first cloud base height below 50~m) or fog (MOR below 500~m) impacts the higher heights. For the data quality we have compared the ZephIR~300 data with that of the reference mast data (up to 200~m) and found very good correlation in the wind speed (slope>0.98 and $r^2>0.99$) and small median deviation in the wind direction (for the four higher height less than 1.5\degree). In about 4\% of the data the wind direction is off by more than 90\degree~($\Delta>90\degree$), which is due to the known 180\degree~issue of the ZephIR~300. This can be significantly reduced by implementing a simple correction scheme on basis of a single reference height. Recently we have relocated the met station from the top of the ZephIR~300 to a separate pole, in order to increase its measuring height from 1.0~m to 1.5~m, and we will investigate whether this reduces the number of $\Delta>90\degree$ events.

The data quality analysis as presented in this paper will be extended to the other outputs of the ZephIR~300, such as the wind gust, standard deviation of horizontal wind speed and the turbulence intensity. Furthermore, the data quality will also be studied as function of the meteorological conditions, i.\,e.\, for the different precipitation, first cloud base height and visibility classes. Also the vertical wind speed output will be investigated, which is more affected by precipitation than the horizontal wind data. The data quality of the ZephIR~300 data at 252~m will be studied by comparing with output from a collocated radar wind profiler. Besides the 10-minute averaged data, we will also investigate the raw data output, which in our configuration is updated every 11~s, whereas the reference mast wind and meteorological data is polled every 12~s.

The motivation to perform this measurement campaign at the CESAR Observatory is the presence of reference wind measurements up to 200~m. However, for the ultimate offshore deployment the meteorological (and atmospheric) conditions are of course different from that of Cabauw. However, by studying the data availability and quality as function of many meteorological quantities, it might be possible to map our results onto its expected conditions offshore. The ZephIR~300 will be placed on the roof deck of the offshore platform, at a height of 42~m above sea level, such that the absolute measuring heights will be higher than in this measurement campaign. Note that on the platform wind measurements (cup anemometer and wind vane) will be present at a height 14~m above the roof deck, making a correction scheme for the 180\degree~error possible.

\section*{Acknowledgment}

We acknowledge Rijkswaterstaat Maritiem Informatievoorziening Service Punt (RWS-MIVSP) for supporting this measurement campaign and providing one of their ZephIR~300 instruments. We also acknowledge Alfons Driever and Corn\'e Oudshoorn (KNMI) for technical support. 

\bibliographystyle{ieeetr}

\bibliography{windlidarbib,cabauwbib}

\begin{thebibliography}{10}

\bibitem{schaderegeling}
``Regulation compensation scheme for offshore grid.'' Regulation of the
  Minister of Economic Affairs of 22 March 2016, no. WJZ/16007215,
  https://offshorewind.rvo.nl/file/download/43059972, p. 102-105.

\bibitem{monna2013}
W.~Monna and F.~Bosveld, ``In higher spheres: 40 years of observations at the
  {Cabauw} site,'' {\em KNMI-Publication 232}, 2013.
\newblock http://bibliotheek.knmi.nl/knmipubmetnummer/knmipub232.pdf.

\bibitem{CESAR}
http://www.cesar-observatory.nl/.

\bibitem{smith2006wle}
D.~A. Smith, M.~Harris, A.~S. Coffey, T.~Mikkelsen, H.~E. J{\o}rgensen,
  J.~Mann, and R.~Danielian, ``Wind lidar evaluation at the {Danish} wind test
  site in {H{\o}vs{\o}re},'' {\em Wind Energy}, vol.~9, p.~87, 2006.

\bibitem{kindler2007aem}
D.~Kindler, A.~Oldroyd, A.~MacAskill, and D.~Finch, ``An eight month test
  campaign of the {Qinetiq} {ZephIR} system: Preliminary results,'' {\em
  Meteorol.\ Z.}, vol.~16, p.~479, 2007.

\bibitem{courtney2008tac}
M.~S. Courtney, R.~Wagner, and P.~Lindel\"ow, ``Testing and comparison of
  lidars for profile and turbulence measurements in wind energy,'' {\em IOP
  Conf.\ Series: Earth and Environmental Science}, vol.~1, p.~012021, 2008.

\bibitem{wouters2016vot}
D.~A.~J. Wouters and J.~W. Wagenaar, ``Verification of the {ZephIR 300} {LiDAR}
  at the {ECN} {LiDAR} {Calibration} {Facility} for the offshore {Europlatform}
  measurement campaign,'' {\em ECN-M--16-029}, 2016.
\newblock https://www.ecn.nl/publications/ECN-E--16-029.

\bibitem{pitter2015itc}
M.~Pitter, C.~Slinger, and M.~Harris, ``Introduction to continuous-wave
  {Doppler} lidar,'' in {\em Remote Sensing for Wind Energy} (A.~{Pe{\~n}a
  \etal}, ed.), ch.~5, p.~99, DTU Wind Energy-E-Report-0084(EN), 2015.

\bibitem{ulden1996abl}
A.~P. {van Ulden} and J.~Wieringa, ``Atmospheric boundary layer research at
  {Cabauw},'' {\em Boundary-Layer Meteorol.}, vol.~78, p.~39, 1996.

\bibitem{verkaik2007wpm}
J.~W. Verkaik and A.~A.~M. Holtslag, ``Wind profiles, momentum fluxes and
  roughness lengths at {Cabauw} revisited,'' {\em Boundary-Layer Meteorol.},
  vol.~122, p.~701, 2007.

\end{thebibliography}
 
\end{document}